\documentclass[12pt]{iopart}

\usepackage{epsfig}

\begin{document}

\letter{Transverse momentum distribution of net baryon number at RHIC}
\author{Steffen A.~Bass\dag\ddag, Berndt M\"uller\dag\ and 
	Dinesh K. Srivastava\S}
\address{\dag\ Department of Physics, Duke University, 
             Durham, North Carolina 27708-0305, USA}
\address{\ddag\ RIKEN BNL Research Center, Brookhaven National Laboratory, 
             Upton, New York 11973, USA}
\address{\S\ Variable Energy Cyclotron 
Centre, 1/AF Bidhan Nagar, Kolkata 700 064, India}            

\ead{bass@phy.duke.edu}

\submitto{\JPG}

\begin{abstract}
We calculate the transverse momentum distribution of net quarks 
(quarks minus antiquarks) in Au+Au collisions at the Relativistic 
Heavy Ion Collider in the framework of the parton cascade model 
at two different rapidities. Parton re-scattering and fragmentation 
is seen to lead to a substantial difference in the slopes of these 
distributions between mid-  and forward-rapidities, in qualitative 
agreement with the corresponding data for the net baryon distribution.
\end{abstract}

\maketitle

Collisions of heavy nuclei at relativistic energies are expected to 
lead to the formation of a deconfined phase of strongly interacting 
nuclear matter, often referred to as a Quark-Gluon-Plasma (QGP). 
Evidences for several of the signatures for the formation of this novel 
state of matter have recently been reported by experiments conducted 
at the Relativistic Heavy Ion Collider (RHIC) at Brookhaven National 
Laboratory \cite{rhic_highlights}. Many aspects of the experimental
data indicate that an equilibrated state of hot and dense matter is
formed in the collisions of Au nuclei at RHIC. One of the open theoretical 
problems is how quickly this thermalized state is formed and which 
mechanisms are responsible for the rapid equilibration. 

It is thus of particular interest to identify processes that can give
information about the pre-equilibrium dynamics in these collisions. 
One possible example of such probes of ``early phase'' physics is the
distribution of net baryon number. Baryon number is a locally 
conserved quantity, and thus its distribution is not easily influenced
by final state interactions. The net baryon number in nucleons is carried
by the valence quarks. Indeed, the valence quark distribution in the
nucleon is {\it defined} as the difference between the quark and antiquark, 
or net quark, distribution. 

Flavour conservation in strong interactions dictates that quarks and 
antiquarks are produced in equal numbers in interactions involving 
sea quarks and gluons, and thus pair production processes do not change 
the net quark distribution. During their passage through dense matter, 
valence quarks collide with other partons and radiate gluons, thereby 
losing some of their longitudinal momentum. In a recent publication
\cite{netb1} we have shown how the valence quark distribution in the 
nucleon, combined with these multiple scattering effects, can explain
the net baryon excess observed in Au+Au collisions at RHIC in the
central rapidity region.

In this letter we shall demonstrate that the net quark distributions
(and thus the measurable net baryon distribution) measured in the final 
state of a heavy ion collision exhibit a strong sensitivity to the dynamics 
of parton transport processes in these collisions. We find that the 
transverse momentum distributions of net quarks vary significantly 
with rapidity, and their broadening at central rapidities is a direct 
measure of the amount of rescattering experienced by the valence quarks. 
Preliminary data reported by the BRAHMS collaboration~\cite{brahms} show 
that the slopes of the net baryon transverse momentum distributions at 
central and forward rapidities differ significantly and thus support 
our findings.

The parton cascade model~\cite{GM92} (PCM) provides a suitable framework 
for the study of the formation of a hot and dense partonic phase, 
starting from clouds of valence quarks, sea quarks, and gluons which 
populate the nuclei. The PCM was devised as a description of the early, 
pre-equilibrium phase of a nucleus-nucleus collision at high energy. 
The current implementation~\cite{VNIBMS} does not include a description 
of the hadronization of the partonic matter and of the subsequent 
scattering among hadrons. These late-stage processes, however, are not 
expected to significantly alter the distribution of net baryon number 
with respect to rapidity, since the net baryon number is locally 
conserved and baryon diffusion in a hadronic gas can be shown to be 
slow \cite{Stephanov}.  No such investigation has been made for the 
transverse momentum distribution of the net baryons, but obviously the 
same arguments apply. Also, we have recently found extensive evidence 
that the transverse momentum distributions of hadrons at RHIC reflect 
the momentum distribution of the partons from which they form~\cite{rainer}. 

Let us briefly recall the fundamental assumptions underlying the PCM.
We assume that the state of the dense partonic system can be 
characterized by a set of one-body distribution functions 
$F_i(x^\mu,p^\alpha)$, where $i$ denotes the flavor index 
($i = g,u,\bar{u},d,\bar{d},\ldots$) and $x^\mu, p^\alpha$ are 
coordinates in the eight-dimensional phase space. The partons are 
assumed to be on their mass shell, except before their first interaction. 
In our numerical implementation, the {\sc GRV-HO} parametrization
\cite{grv} is used, and the parton distribution functions are sampled 
at an initialization scale $Q_0^2$ to create a discrete set of particles. 
Partons generally propagate on-shell and along straight-line trajectories 
between interactions. Before their first collision, all partons move with 
the beam (target) rapidity and do not have an ``intrinsic'' transverse 
momentum.

The time-evolution of the parton distribution is governed by a 
relativistic Boltzmann equation:
\begin{equation}
p^\mu \frac{\partial}{\partial x^\mu} F_i(x,\vec p) = {\cal C}_i[F]
\label{eq03}
\end{equation}
where the collision term ${\cal C}_i$ is a nonlinear functional 
of the phase-space distribution function. The calculations discussed
below include all lowest-order QCD scattering processes between 
massless quarks and gluons. A low momentum transfer cut-off 
$p_T^{{\rm min}}$ is needed to regularize the infrared divergence 
of the perturbative parton-parton cross sections.  Additionally, we 
include the branchings $q \to q g$, $q \to q\gamma$, $ g \to gg$ and 
$g \to q\overline{q}$ \cite{frag}.  The soft and collinear singularities 
in the showers are avoided by terminating the branchings when the 
virtuality of the time-like partons drops below $\mu_0 = 1$ GeV. 
The results to be discussed below have been obtained using the 
{\sc VNI/BMS}~\cite{VNIBMS} implementation of the PCM.

\begin{figure}[tb]   
\centerline{\epsfig{file=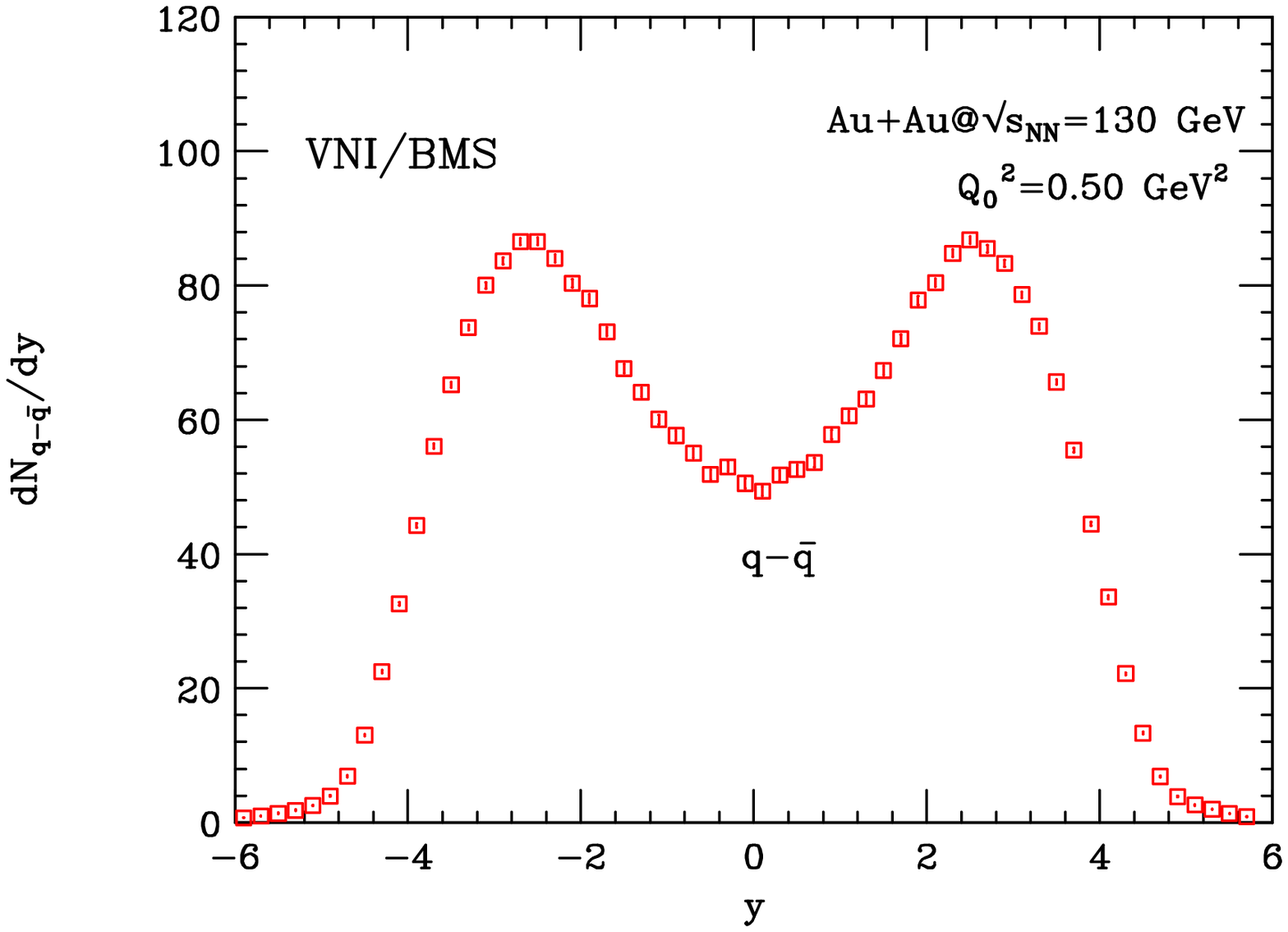,width=8cm}
\epsfig{file=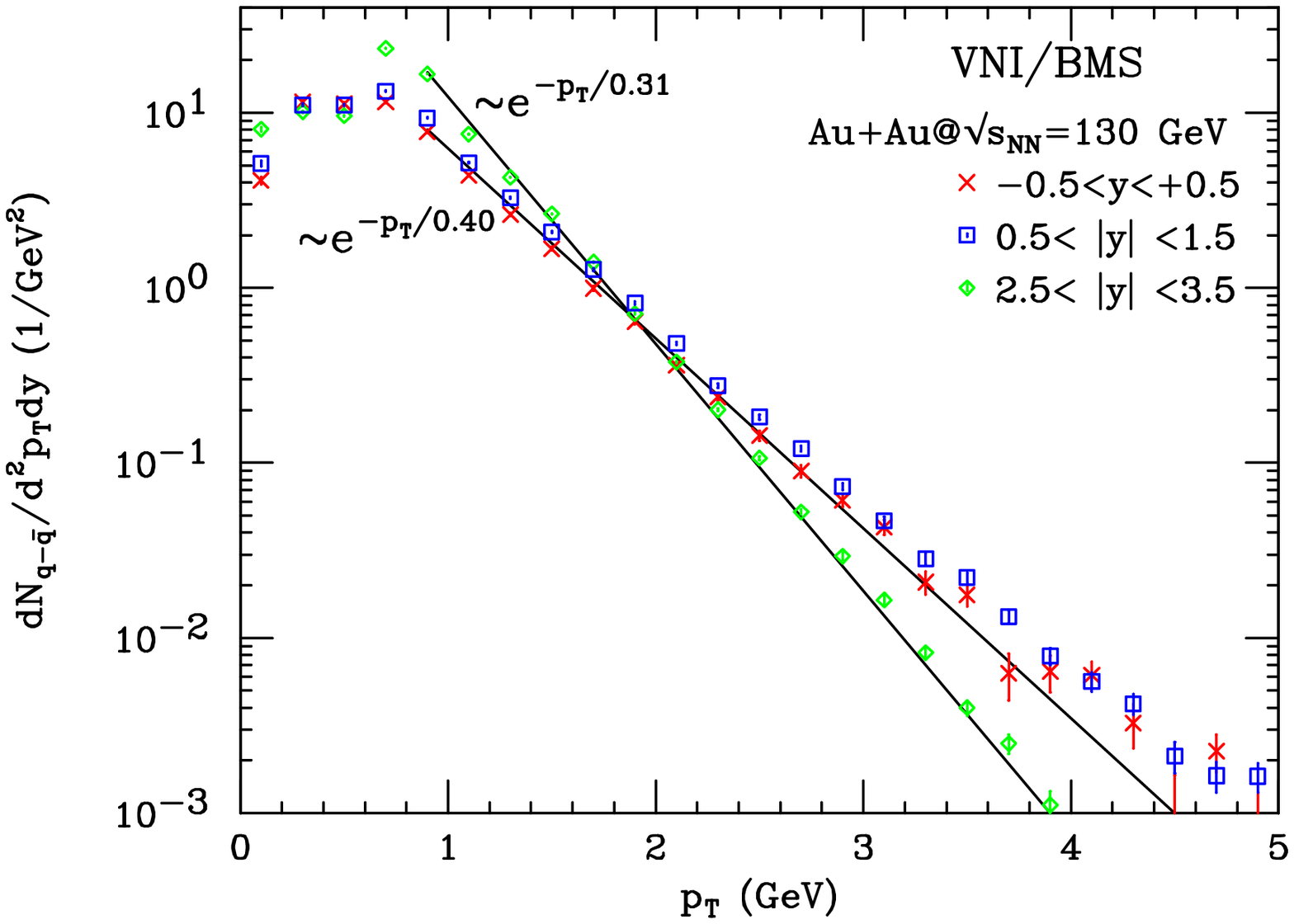,width=8cm}}
\caption{Net baryon number rapidity distributions (left) and $p_T$ 
distributions at different rapidities (right) for Au+Au reactions at
$\sqrt{s}_{NN}=130$~GeV in the parton cascade model VNI/BMS. The 
estimated slopes for the $p_T$ spectra are shown. Multiple collisions 
among partons as well as parton multiplication due to time-like 
branchings are included. The curves show the estimated slopes.}
\label{fig1}
\end{figure}

We shall start our investigation by exploring the transverse momentum 
distribution of the net quarks as a function of rapidity. The left 
frame of Fig.~\ref{fig1} shows the net quark rapidity distribution 
for a Au+Au collision at $\sqrt{s}_{NN}=130$~GeV, using the
initialization scale and low momentum cut-off scales of the pQCD cross 
sections, $Q_0^2 = (p_T^{{\rm min}})^2 = 0.50$~GeV$^2$. The calculations 
included multiple scatterings among the partons as well as their 
fragmentation by time-like branchings. We have already shown~\cite{netb1}
that the corresponding value for the net baryons is in good agreement 
with the results obtained by PHENIX and STAR experiments~\cite{bbar_exp}.

\begin{figure}[tb]   
\centerline{\epsfig{file=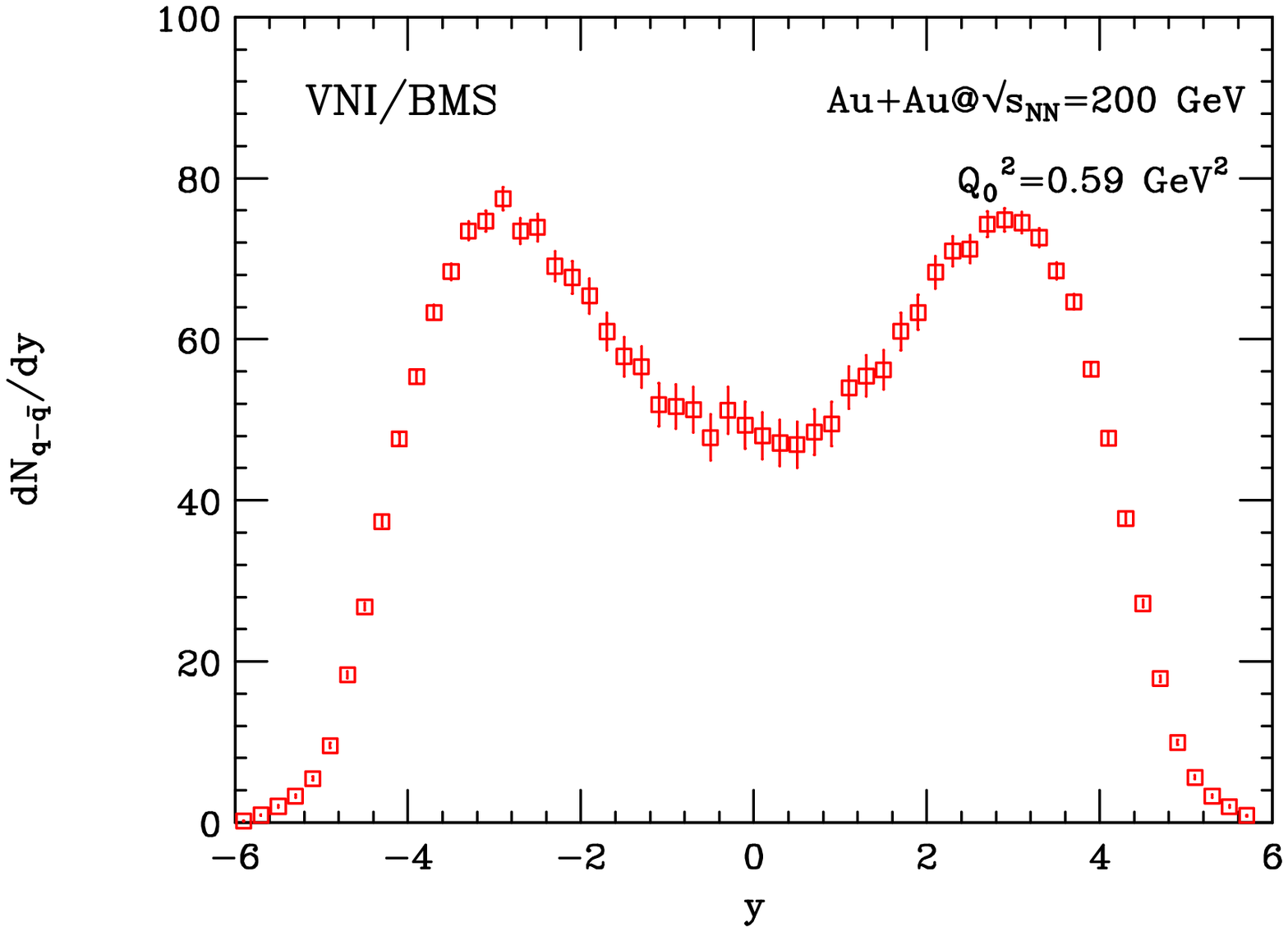,width=8cm}
\epsfig{file=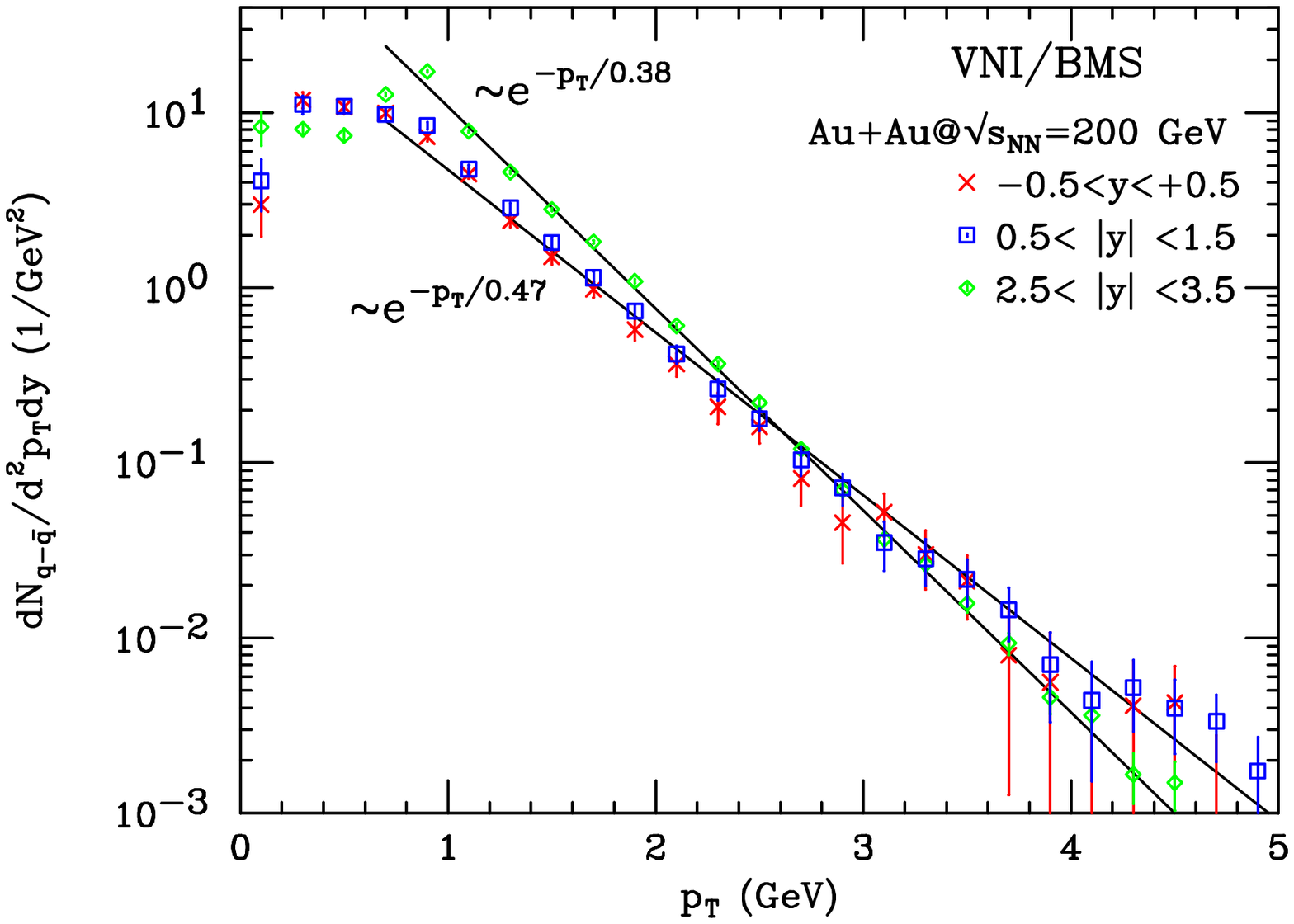,width=8cm}}
\caption{Same as Figure~\ref{fig1} for $\sqrt{s}_{NN}=200$~GeV.}
\label{fig2}
\end{figure}

We now focus on the transverse momentum distribution of the net quarks 
(right panel). We see that the spectral slopes are essentially 
identical for $y_{\rm CM}=0$ and $y_{\rm CM}=1$ but significantly steeper 
for $y_{\rm CM}=3$. We also find that for $p_T > p_T^{{\rm min}}$ the 
spectra are well represented by exponentials, with the ``temperature''
at mid-rapidity being about 20\% larger than at more forward rapidity.
The corresponding results for $\sqrt{s}_{NN}=200$~GeV are given in 
Fig.~\ref{fig2} and confirm our findings for $\sqrt{s}_{NN}=130$~GeV.
If the slopes of the momentum distributions of hadrons remain proportional 
to the slopes for the partons as predicted by the recombination model 
\cite{rainer}, the differences in the slopes of net baryons at $y_{\rm CM}=0$ 
and $y_{\rm CM}=3$ will reflect this difference in agreement with the preliminary
experimental findings.

\begin{figure}[tb]   
\centerline{\epsfig{file=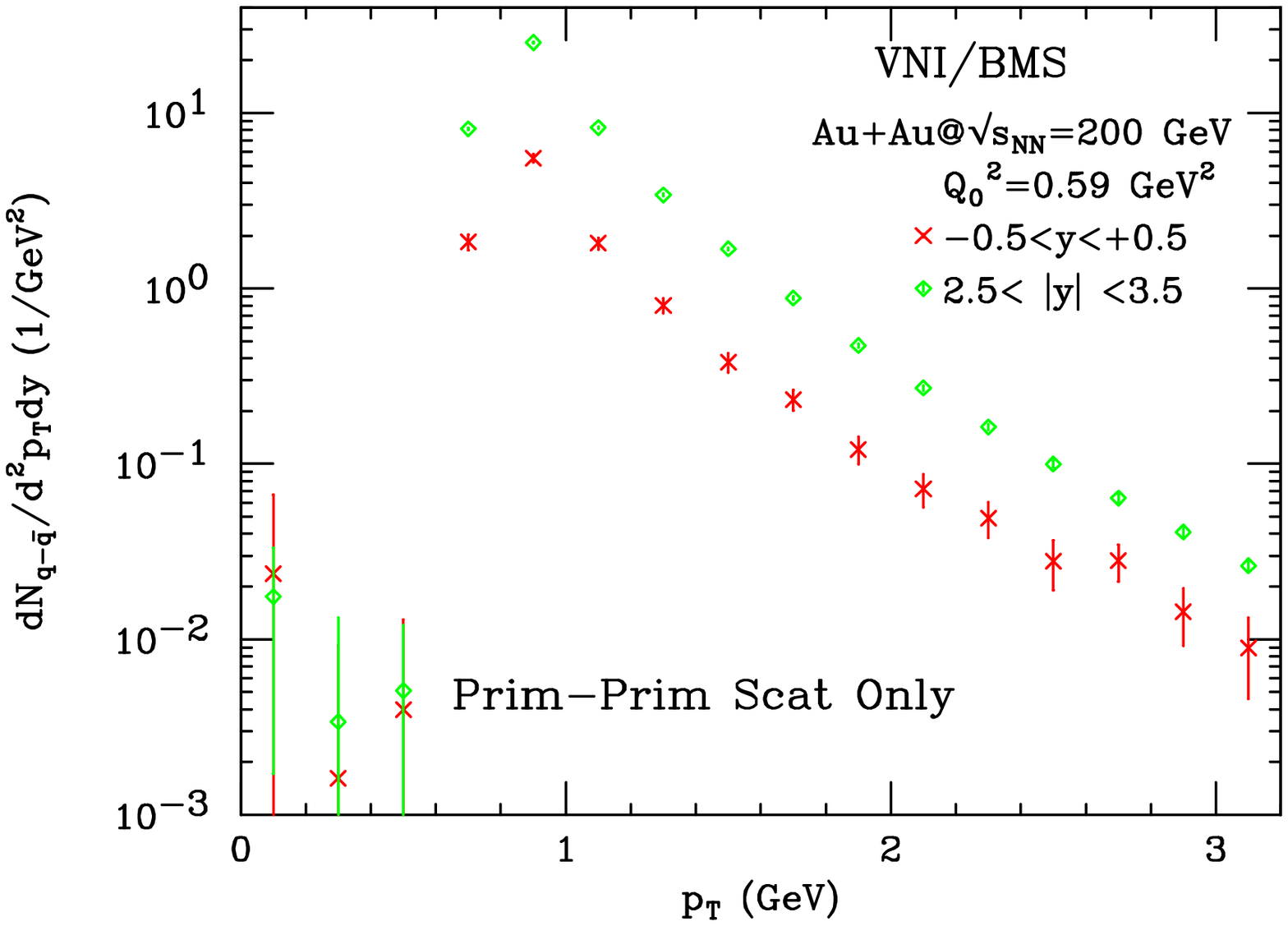,width=8cm}
\epsfig{file=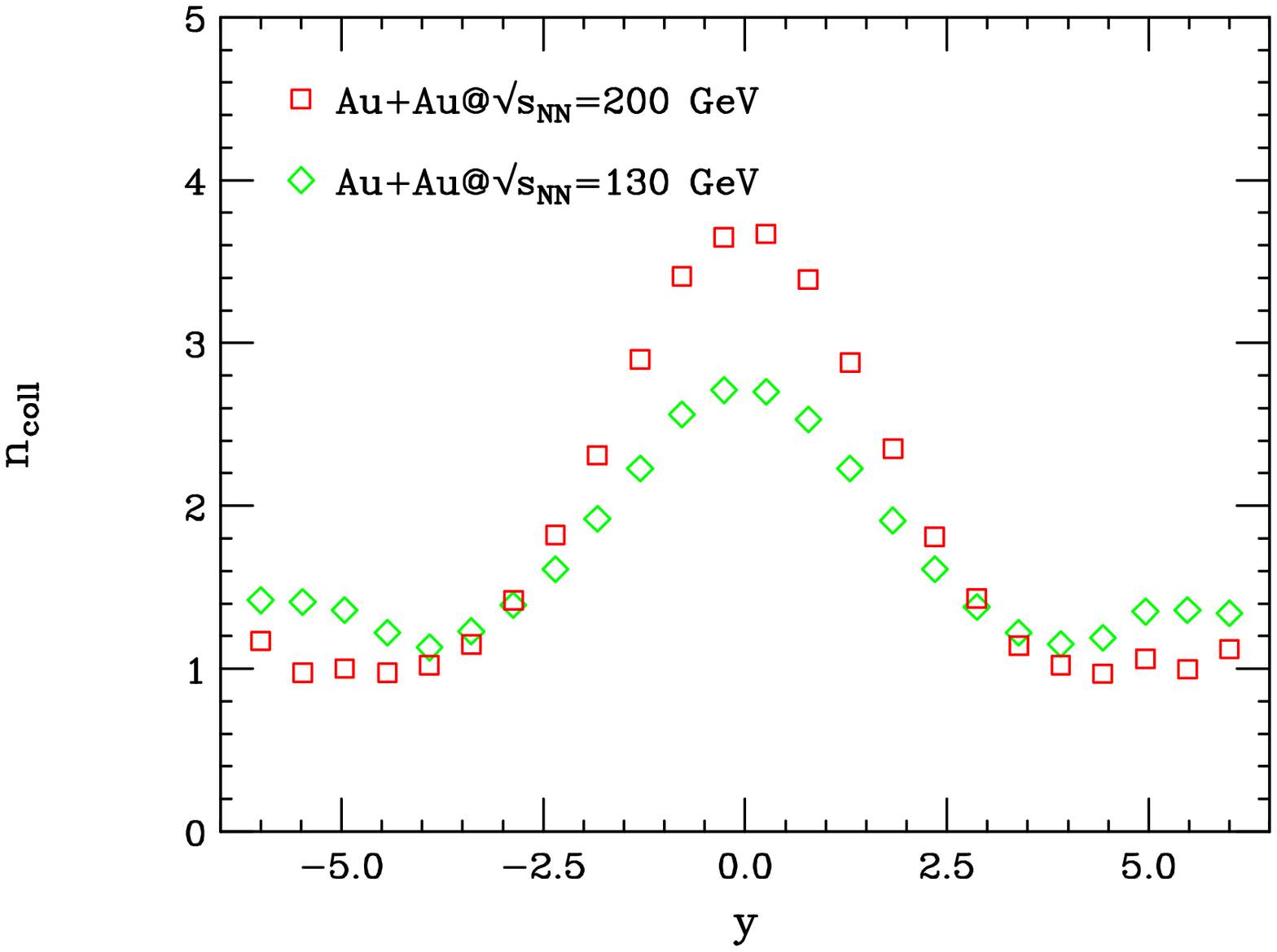,width=8cm}}
\caption{
Left frame: transverse momentum distribution of net quarks
 at different rapidities (bottom) for Au+Au reactions at
$\sqrt{s}_{NN}=200$~GeV, using PCM when only primary-primary
collisions are included. 
Right frame: The rapidity distributions  of number of collisions 
suffered by partons at $\sqrt{s}_{NN}=130$ and $200$ GeV,  
in Au+Au collisions using PCM. Both multiple scatterings and 
time-like branching of the partons are included.}
\label{fig3}
\end{figure}

The rapidity dependence of the transverse momentum slopes may be caused
by a combination of two effects:
\begin{enumerate}
\item Initial state rapidity--$p_T$ correlations: as we have shown in a 
previous publication \cite{netb1}, initial (primary-primary) parton 
scattering ``releases'' valence quarks at a rapidity corresponding to the 
Bjorken-$x$ they carry according to the parton distribution function
of their mother hadron. This {\em predetermined} rapidity (and thus 
longitudinal momentum and energy) may cause the associated transverse 
momentum distribution in the primary-primary scattering to become a 
function of rapidity as well. 
\item Parton-parton rescattering: the rapidity dependence of the slopes
may be due to partons rescattering more often around mid-rapidity than 
at forward rapidities.
\end{enumerate}
In order to distinguish between these two mechanisms (or to determine 
their relative importance) we perform two analyses.
First, we perform a PCM calculation for a Au+Au collision at 
$\sqrt{s}_{NN}=$ 200 GeV, in which we restrict the interactions among
the partons to primary-primary collisions, i.e. allowing each parton to 
scatter only once. The results are shown in the left-hand frame of 
Fig.~\ref{fig3}. The spectra at $y_{\rm CM}=0$ and $y_{\rm CM}=3$ 
have nearly identical shapes and both exhibit a power-law behaviour, 
as expected from pQCD. From this calculation we may conclude that the 
observed rapidity-dependence of the transverse momentum slopes is not 
caused by rapidity--$p_T$ correlations in the initial state.

On the other hand, when we plot the average number of
collisions suffered by partons as a function of their rapidity 
(right frame of Fig.~\ref{fig3}), we find that the partons ending 
up at the central rapidities suffer on an average 
many more collisions than those which end up at larger rapidities. 
This result provides a strong indication that the rapidity
dependence of the spectral slopes is directly related to the 
amount of rescattering experienced by the partons, and that
an exponential shape of the parton transverse momentum distribution 
has its origin in the multiple interactions included in the PCM.
In addition, we find that the shape of the $n_{\rm coll}$ vs. $y$ 
distribution strongly resembles the shape of the rapidity 
distribution of direct photons produced in parton-parton scatterings, 
which we have previously shown to be proportional to the number of 
parton interactions occurring during the evolution of the heavy-ion 
collision \cite{bms_photon}.

\begin{figure}[tb]   
\centerline{\epsfig{file=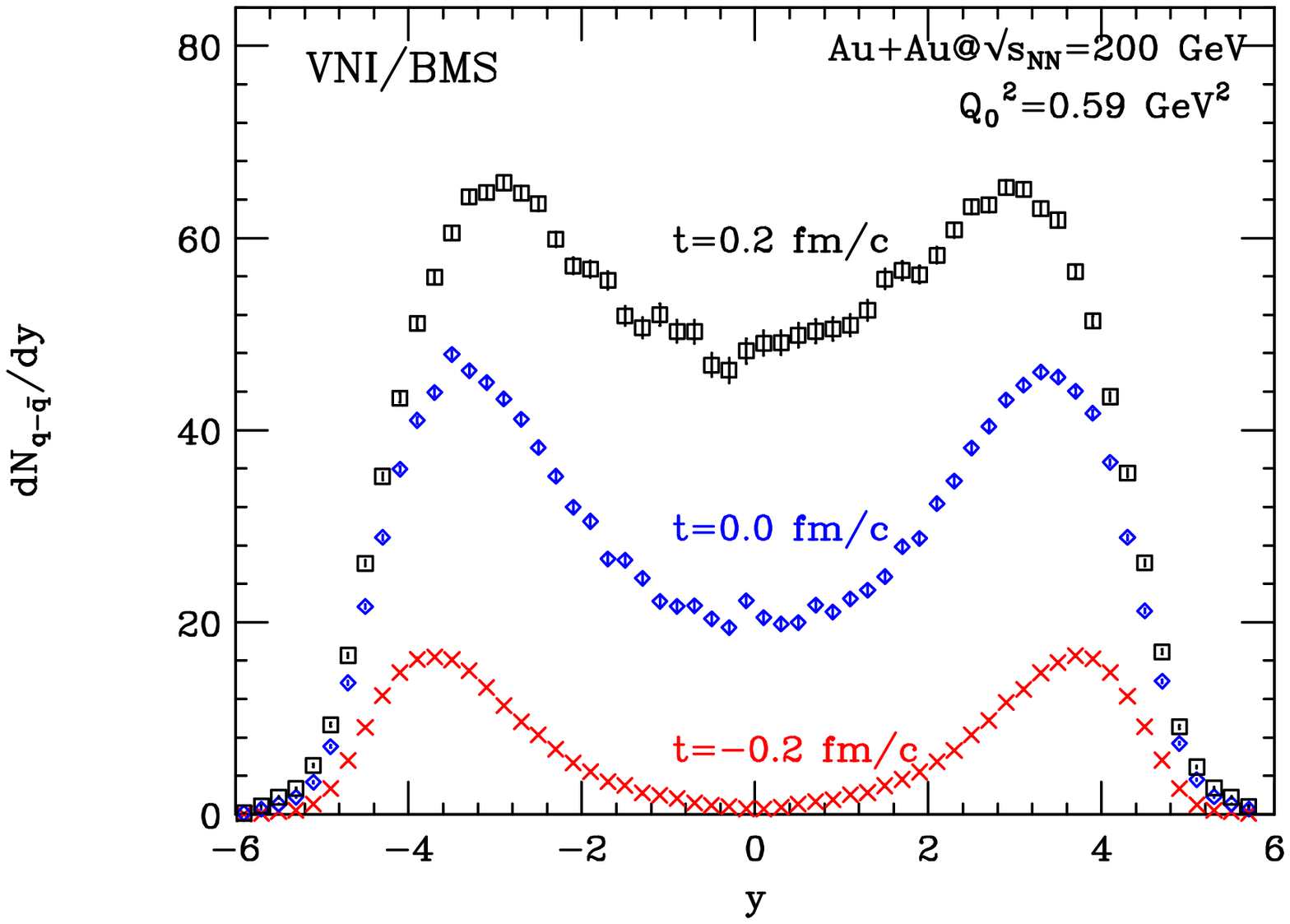,width=8.6cm}}
\centerline{\epsfig{file=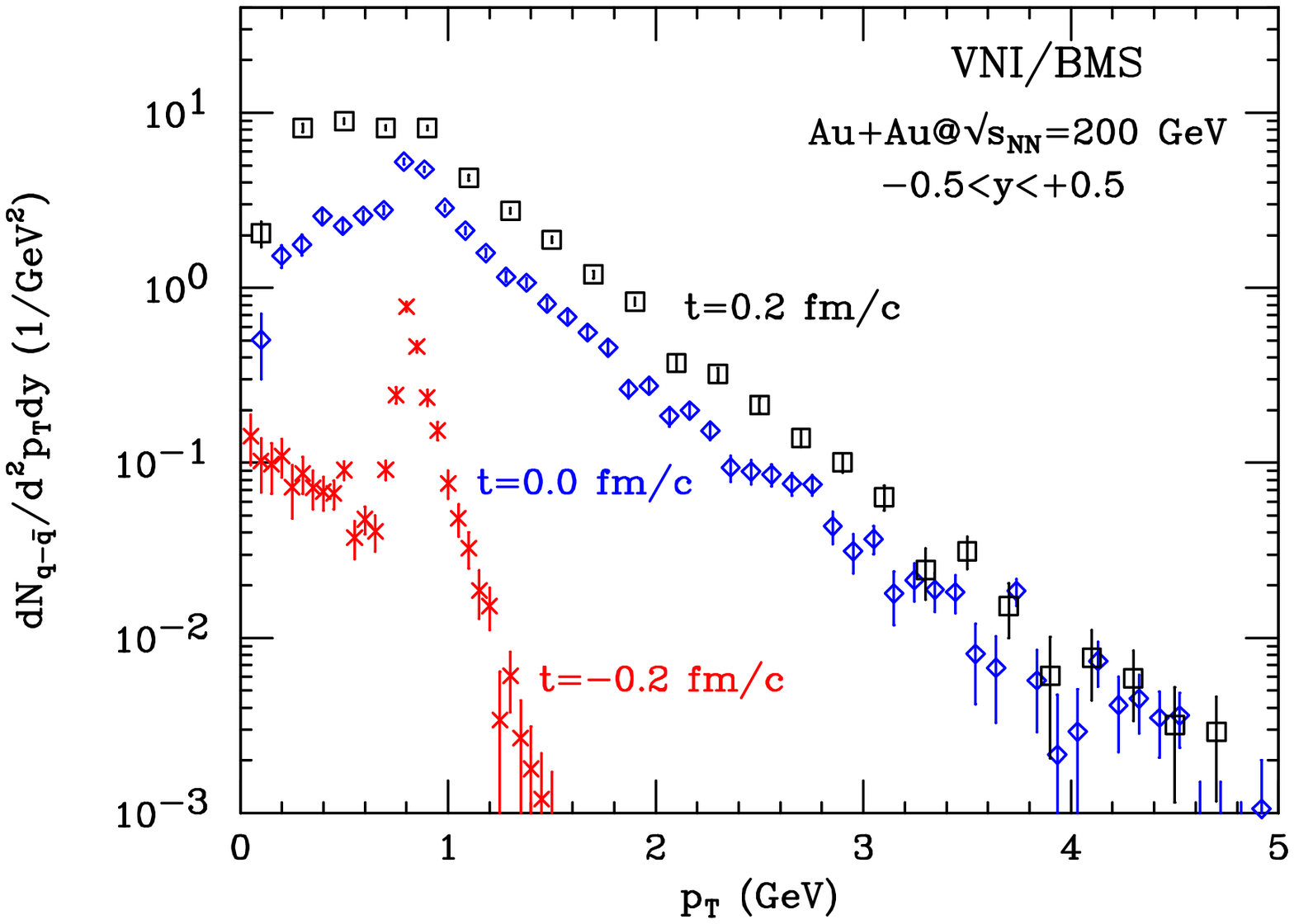,width=8.6cm}}
\centerline{\epsfig{file=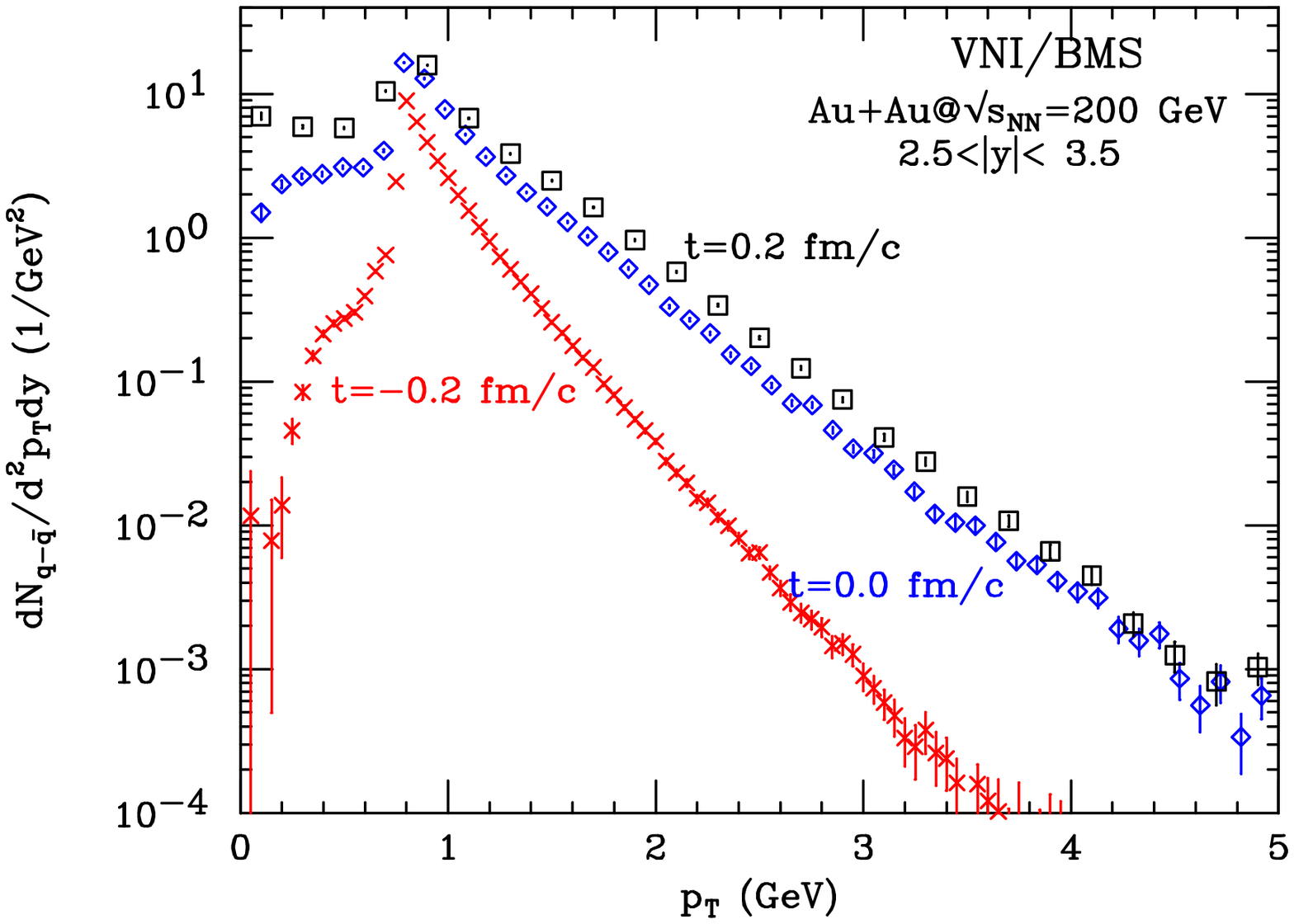,width=8.6cm}}
\caption{Rapidity distributions (top) and $p_T$ distribution at  
$y_{\rm CM}=0$ (middle) and $y_{\rm CM}=3$ (bottom) for net-quarks at t=-0.2 fm/$c$, 
0.0 fm/$c$, and 0.2 fm/$c$ in Au+Au  collisions at 
$\sqrt{s}_{NN}=200$~GeV, as predicted by VNI/BMS.}
\label{fig5}
\end{figure}

The PCM allows us to study the full space-time evolution of the parton 
distributions. We therefore are able to trace the evolution of the net 
quark momentum distributions at different rapidities from their initial 
release via primary-primary scattering to their final shape.
Figure~\ref{fig5} shows the net quark rapidity distribution (top frame)
as well as the net quark transverse momentum distributions at 
$y_{\rm CM}=0$ (middle frame) and $y_{\rm CM}=3$ (bottom frame) at times
t=-0.2 fm/$c$, 0 fm/$c$, and 0.2 fm/$c$, where $t=0$ fm/$c$ 
corresponds to the time of maximal overlap of the two colliding nuclei.
Note that the longitudinal spread of the partons according to
$\Delta z \sim 1/p_z$ around the centre of the nucleus, incorporated in
the PCM, permits collisions between them even when the nuclei have not
yet obtained full overlap, provided the collisions are permitted 
by the cut-offs implemented in the PCM. 

Several interesting observations emerge from this analysis: The momentum 
distribution of the net quarks at larger rapidities gets frozen fairly 
early during the collision, while the corresponding distributions at 
central rapidities continue to evolve. This is true for both the
rapidity distributions as well as the transverse momentum distributions.

This observation suggests that the net quark (or the net baryon) 
distribution at different rapidities can serve as a useful tool
to explore the interactions among partons (quarks) during the 
early phase of the heavy-ion collision. For example, the study of 
matter emitted at larger rapidities will offer insight into the 
very early stage of a relativisitic heavy-ion reaction at which the
initial parton distributions decohere (and at which saturation phenomena 
\cite{colorglass} may be of importance), whereas the mid-rapidy domain 
will provide information about a system of strongly and multiply 
interacting partons with all its associated phemonena, ranging from 
parton equilibration to jet energy-loss. 

In brief, we have studied the rapidity dependence of the net quark
distribution in relativistic heavy ion collisions using the parton 
cascade model. We find that the inverse slopes of the transverse momentum 
distributions at $y_{\rm CM}=0$ and $y_{\rm CM}=3$ in Au+Au collisions 
at $\sqrt{s}_{NN}=130$ and 200 GeV differ by about 20\%, in agreement 
with recently reported measurements by BRAHMS \cite{brahms}.  Our 
analysis shows that this difference arises due to the different amount 
of multiple scatterings suffered by partons at central and forward 
rapidities. 

\ack  
This work was supported in part by RIKEN, the Brookhaven National 
Laboratory, and DOE grants DE-FG02-96ER40945 and DE-AC02-98CH10886.

\end{document}